%% file: PIMRC.tex
\documentclass[conference]{IEEEtran}
\usepackage[letterpaper, left=1in, right=1in, bottom=1in, top=0.75in]{geometry}

\usepackage{amsfonts}
\usepackage{etex}
\usepackage{amsmath,amssymb}
\usepackage{mathtools,hyperref}
\usepackage{amsthm}
\usepackage{float}
\usepackage{slashbox}
\usepackage{subcaption}
\usepackage{nicefrac}
\usepackage[dvipsnames]{xcolor}
\usepackage{tikz}
\usepackage{ctable}
\usepackage{graphicx}  
\usepackage{tabularx}
\usepackage{accents}
\usepackage{enumitem}
\usepackage[makeroom]{cancel}
\usepackage{colortbl}
\usepackage{multirow,balance}
\usepackage{mathtools}
\usepackage{pgfplots}
\pgfplotsset{compat=newest}
\usetikzlibrary{plotmarks}
\usepackage{grffile}
\usepackage{amsmath}

\usepackage{balance}
 
\allowdisplaybreaks 

\newcommand{\argmax}{\arg\!\max}
\begin{document}

\title{Waveform and Spectrum Management for Unmanned Aerial Systems Beyond 2025}

\author{\IEEEauthorblockN{Jaber Kakar}
\IEEEauthorblockA{Institute of Digital Communication Systems\\
Ruhr-Universit{\"a}t Bochum\\
Email: jaber.kakar@rub.de}
\and
\IEEEauthorblockN{Vuk Marojevic}
\IEEEauthorblockA{Bradley Dept. of Electrical and Computer Engineering\\
	Virginia Tech\\
	Email: maroje@vt.edu}
}

\maketitle

\begin{abstract}
The application domains of civilian unmanned aerial systems (UASs) include agriculture, exploration, transportation, and entertainment. The expected growth of the UAS industry brings along new challenges: Unmanned aerial vehicle (UAV) flight control signaling requires low throughput, but extremely high reliability, whereas the data rate for payload data can be significant. This paper develops UAV number projections and concludes that small and micro UAVs will dominate the US airspace with accelerated growth between 2028 and 2032. We analyze the orthogonal frequency division multiplexing (OFDM) waveform because it can provide the much needed flexibility, spectral efficiency, and, potentially, reliability and derive suitable OFDM waveform parameters as a function of UAV flight characteristics. OFDM also lends itself to agile spectrum access. Based on our UAV growth predictions, we conclude that dynamic spectrum access is needed and discuss the applicability of spectrum sharing techniques for future UAS communications.
\end{abstract} \begin{IEEEkeywords}
UAV, spectrum sharing, OFDM.
\end{IEEEkeywords}

\IEEEpeerreviewmaketitle

\input{content/introduction}   
\input{content/prediction} 
\input{content/a2g}

\input{content/waveform}
\input{content/spectrum_sharing}

\bibliographystyle{IEEEtran}
\bibliography{content/bibliography}
\balance
 
\end{document}

%% file: content/introduction.tex
\section{Introduction}
\label{sec:intro}

Unmanned aerial systems (UASs) already outnumber
traditional manned aircraft (AC) systems for military missions. Unmanned aerial vehicle (UAV) and system technology is fairly advanced and development as well as maintenance costs are significantly lower for unmanned than for manned AC systems \cite{Austin}. Not only are UAVs cost effective, but the applications for government and commercial
purposes are abundant. UAVs are used for transportation of goods, for supporting or replacing wireless communications infrastructure, and for supporting agriculture, security and entertainment industries, among
others \cite{ReportITU}. For instance, as part of Google Project Loon, high
altitude UAV Internet access points (APs) were proposed as alternatives
for terrestrial APs. Qualcomm recently completed its testing with a major US carrier of UAV communications through the 4G LTE cellular infrastructure.

Small and micro UAVs (SUAV/MAV) are low-altitude UAV alternatives
that are suitable for dense urban scenarios. 
Future SUAVs and MAVs will fall under the Internet of Things (IoT) umbrella and provide services such as environmental sensing. Other use cases are communications relaying, broadcasting, and radio frequency (RF) spectrum sensing. Recent predictions, conducted by the US National Transportation Center, reveal that the number of UAVs for commercial purposes will outnumber UAVs owned by DoD by a factor of 10 or more by 2035. SUAVs and MAVs of
less then ten feet in size and under fifty-five pounds in weight will likely dominate the US
airspace. The radio communications links of these ACs will be line of sight (LoS) for typical deployments because of their limited range.

It is expected that the advances in UAS
technology and benefits of UASs for commercial and other civilian
operations will bring along new challenges: safety and efficiency of operation
through real-time exchange of (1) latency-sensitive control data and (2) throughput-intensive payload data that are captured by the UAV sensors. These challenges reduce to one of the fundamental problems in modern wireless communications: spectrum management.
UAV spectrum in the 1755 MHz band, which is part of the AWS-3 band, is considered
for relocation in the US. Until this happens, secondary users (commercial LTE systems) will need to coexist with legacy users (military UASs). The AWS-3 band was auctioned in the US in 2015 \cite{ntia} and research is underway for making harmonious coexistence possible. 

The exchange of rich content data or streaming high-definition video 
requires a significant amount of spectrum, proportional to
the desired throughput and quality. When the air becomes more
congested with UAVs and the airwaves with RF signals, a significant amount of bandwidth will need to be provisioned to accommodate the desired communications needs.

The ITU bandwidth calculations are rather pessimistic
because they only account for time-sparse data exchange
for sense and avoid (S\&A) applications in environments with
relatively low UAV densities. We believe that future air-to-ground
(ATG) links will be (1) interference-limited and (2) throughput-intensive and that enough
dedicated spectrum will not be available. As a result, RF spectrum needs to be managed differently, avoiding excessive interference and providing enough resources for efficient communications for a growing number or spectrum consumers.
Sharing spectrum dynamically enables providing bandwidth on demand, when and where needed, and is a feature of 5G and emerging 4G LTE systems. 

Research on UAS spectrum sharing has been scarce.
Brown et al. \cite{Brown, Brown2} suggest the deployment of a
policy-based cognitive radio for UAV spectrum sharing and 
discusses adaptation of policy-based radios.
In \cite{McHenry}, dynamic spectrum access (DSA) is proposed as the solution to the spectrum crunch for military UAS. 

This paper first projects the bandwidth requirement for future UASs as a function of UAV growth projections and new predictive models (Section II). We then analyze the suitability of the orthogonal frequency division multiplexing (OFDM) waveform, which is used in LTE and WiFi, for UAS communications, because of the huge R\&D support and inherent flexibility and adaptability to operate in unlicensed and shared spectrum (Section III and IV). Without loss of generality, our analysis focuses on direct
ATG LoS links. 
We finally discuss the suitability of emerging spectrum sharing technologies and the integration of terrestrial and UAS communications networks. 

%% file: content/prediction.tex
\section{Projection of UAS Growth and Spectrum Requirements} 

Predicting the numbers of UAVs is an important step to determine RF
spectrum requirements. A UAS consists of a ground control
station (GCS) and one or several UAVs. To address the spectrum
requirements for control and non-payload communications
(CNPC), ITU and NASA have conducted projections on the evolution
of UAVs \cite{ReportITU, NASA}. We believe that the figures are conservative and use \cite{Volpe,DoD} to quantify the UAV-type specific numbers from 2015 until 2035. Reference \cite{DoD} predicts future demand of UAVs for DoD, public safety and the commercial sectors. Reference \cite{Volpe} suggests
an s-curve shaped functional relationship for characterizing the
number of UAVs between 2015 and 2035 for the commercial and
public sectors. Hence, we estimate the number of UAV
until 2035 using
\begin{equation}\label{eq:demand} f(x)=p_{1}+\frac{(p_{2}-p_{1})}{1+10^{p_{4}(p_{3}-x)}},\end{equation}
where $x=t_{year}-2015$. The curve fitting results for commercial UAS and total public agencies (including DoD) are shown in Table \ref{tab:fitting}. 

\begin{figure}[h]
  \centering
{\includegraphics[width=3.0in]{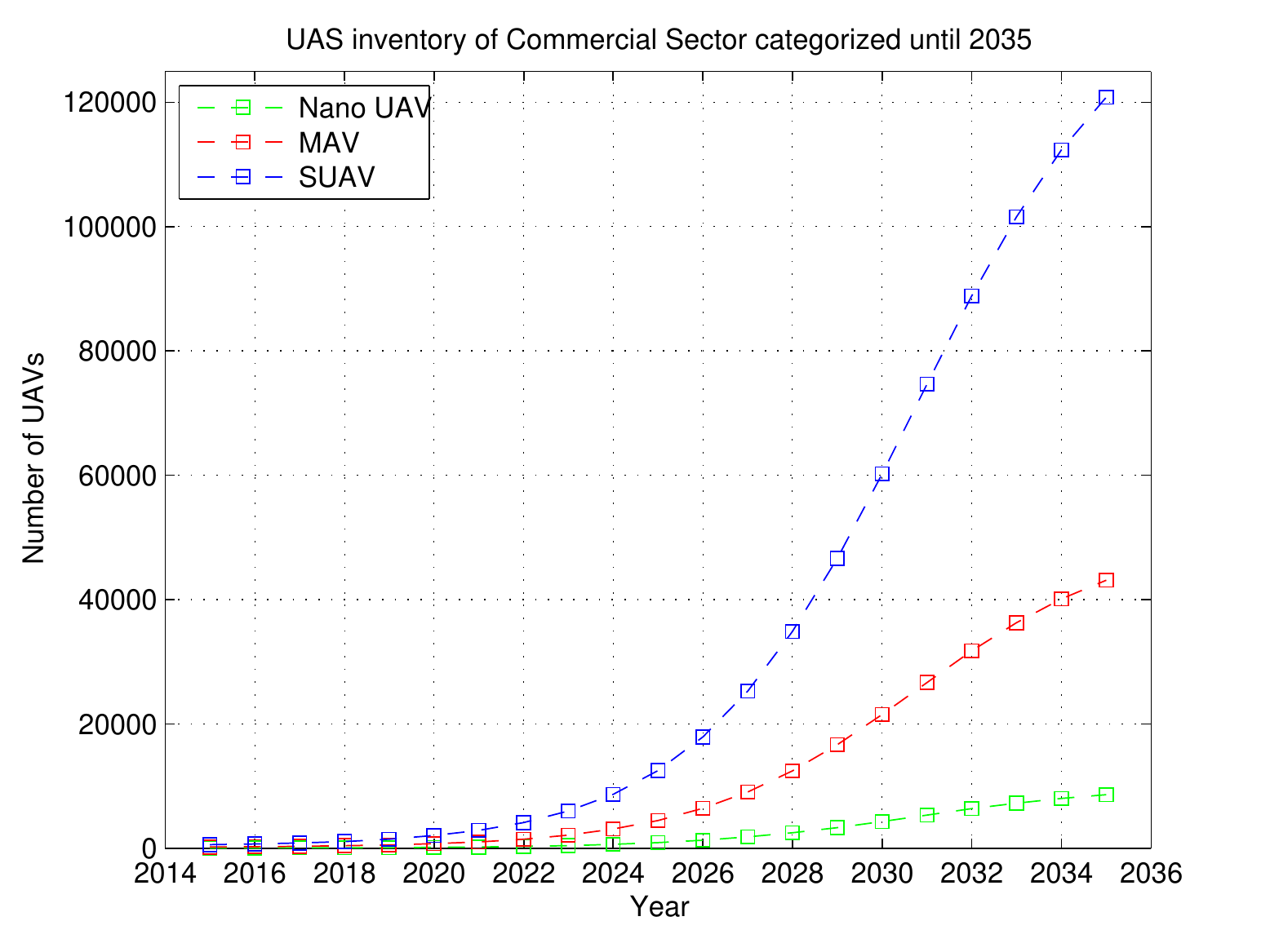}}
  \caption{\label{fig:b}\small Quantites of commercial UAVs.}
\end{figure}

\begin{table}[h]
\begin{center}
\footnotesize
\caption{\small Projected UAS numbers until 2035 based on \cite{Volpe}}
\label{tab:fitting}
\scalebox{0.95}{
\begin{tabular}{|c|c|c|c|c|} \hline
& $p_{1}$ & $p_{2}$ & $p_{3}$ & $p_{4}$ \\ \hline
Commmercial & 487.95 & 2.03$\cdot 10^{5}$ & 15.75 & 0.18 \\ \hline
Federal Agencies & 207.22 & 1.02$\cdot 10^{4}$ & 9.73 & 0.18 \\ \hline
State and Local Agencies & 1.87$\cdot 10^{3}$ & 4.64$\cdot 10^{4}$ & 12.49 & 0.19 \\ \hline
\end{tabular}
}
\end{center}
\end{table}

\begin{table*}[t]
\footnotesize
\begin{center}
\caption{\small Estimation of average UAV densities in 2030.}
\label{tab:density}
\begin{tabular}{ccc|c|c|c|}
\cline{4-6}
& & & \multicolumn{3}{c|}{UAV} \\ \cline{4-6}
& & & \multicolumn{1}{c|}{Small} & Medium & Large \\ \hline
\multicolumn{3}{|c|}{Effective Number of UAVs in operation by 2030} & 7,229 & 8,919 & 760 \\ \hline
\multicolumn{1}{|c}{\multirow{3}{*}{UAV Density [UAV/10000 km$^{2}$]}} & \multicolumn{1}{|c|}{Low Altitude} & $<1500$ m & 7.33 & -- & -- \\ \cline{2-6}
 \multicolumn{1}{|c|}{} & \multicolumn{1}{c|}{Medium Altitude} & $>1500$ m and $<6000$ m & -- & 9.05 & -- \\ \cline{2-6}
 \multicolumn{1}{|c|}{} & \multicolumn{1}{c|}{High Altitude} & $>6000$ m & -- & -- & 0.77 \\ \hline
\end{tabular}
\end{center}
\end{table*}

\begin{figure}[h]
\includegraphics[width=3.3in]{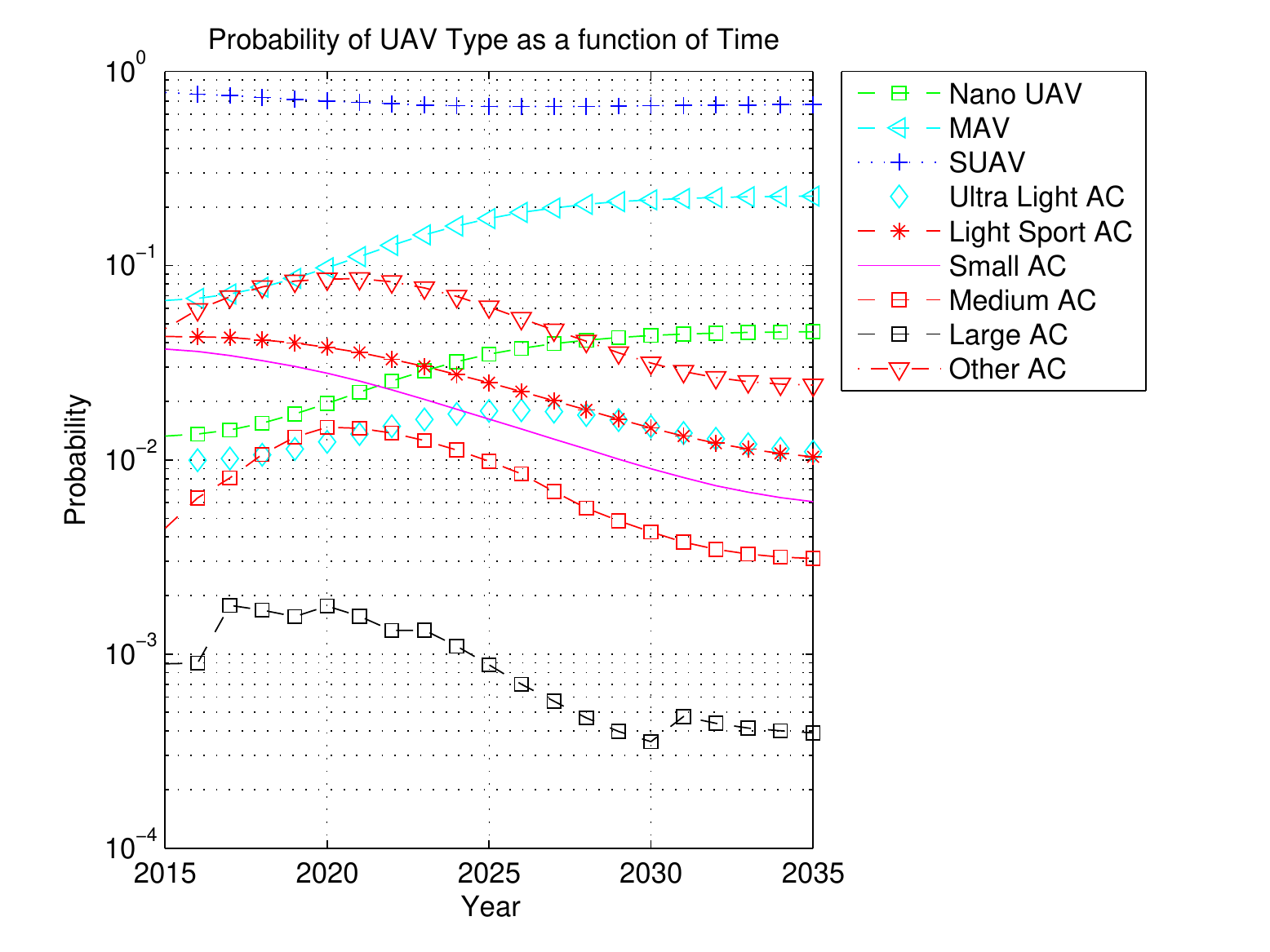}
\caption{\label{fig:results_prob}\small UAV type probability mass function over time.}
\end{figure}

Our analysis indicates that commercial UAVs will outnumber
public agency UAVs. (The DoD expects a linear increase
of their UAV fleets, which will be outnumbered by commercial
UAVs within the next ten years.) Figure \ref{fig:b} shows the evolution of Nano UAV, MAV and SUAV numbers 
for commercial purposes. We use the total numbers (including estimates for DoD-owned UAVs) to determine
the probability mass function (pmf) of UAV types, distinguishing among Nano UAVs, MAVs, SUAVs, Ultralight ACs, Light Sport ACs, Small ACs, and Medium ACs. The
time-dependent pmf for 2015-2035 can be seen in Figure \ref{fig:results_prob}. It
shows that SUAVs and MAVs are expected to dominate the UAV
market because of their low cost and versatility.

In order to estimate the bandwidth requirements for CNPC in 2030,
we use ITU’s methodology 1 \cite{ReportITU}. Table
2 of \cite{ReportITU} specifies the data rate requirements for command
\& control (C2), air traffic control (ATC) relay, S\&A (including video and weather radar data) as a function of the UAV altitude. C2 links include navigational information
and telecommands in the uplink (UL) and telemetry and
navigational display data in the downlink (DL). 
S\&A data mainly contains target tracking (3D position, velocity,
timestamp, etc.), weather radar and non-payload video data
for temporary awareness of the environment.

We use our estimates for commercial and public agency UAVs
for 2030 according to Fig. \ref{fig:b} to determine the CNPC bandwidth
requirements for LoS communications. We assume that
around 88\% of commercial UAVs in Fig. \ref{fig:b} belong to the
agricultural sector \cite{Volpe}. This percentage is not considered for
CNPC bandwidth computation. Furthermore, we assume that
about 15\% of all public UAVs will be used on regular basis.
Using the probability of Fig. \ref{fig:results_prob} and the typical altitude of
operation of small, medium and large UAV, we can calculate
the altitude-specific UAV densities (using the US total area of
around $9.8$ million $\text{km}^{2}$) \cite{ReportITU}. The results are shown in Table \ref{tab:density}. Note that small, medium and large UAVs in Table \ref{tab:density} are used
to classify the altitude of operation rather than the UAV type. 
These UAV density numbers differ from ITU's
results by a factor of approximately $1.2$ for small and large
and $5.8$ for medium UAVs.

The determined densities are used to compute the number of
UAVs per cell. ITU defines cell types A, B, C and D 
for terrestrial communications to accommodate UAVs
with different operational altitudes. Using the exact same link
and cell configurations as in \cite{ReportITU}, our CNPC bandwidth estimates
are 69.5 and 39.5 MHz for a terrestrial communications infrastructure
with and without video and weather radar data. For comparison, ITU’s values are 33.9
and 15.9 MHz, respectively. It is interesting to mention that \cite{ReportITU}
considers a spectral efficiency of 0.75 bps/Hz for all CNPC links.

In order to satisfy the CNPC bandwidth requirements (including
video and weather radar data) according to our estimates with the
designated bandwidth of 34 MHz, the spectral efficiency
needs to improve to about 1.53 bps/Hz. The payload data typically requires much higher data rates and the amount of bandwidth cannot be accurately estimated, but will extend previous MHz estimates by orders of magnitude.

%% file: content/a2g.tex
\section{ATG Communications Channel} 

The simplest UAS in the segregated airspace
consists of the UAV and the ground control station (GCS) with
exclusive frequency assignment for the UAV-GCS (DL) and GCS-UAV (UL) links. Based on the actual flight range, we need to differentiate between LoS and beyond LoS (BLoS) communications. In this paper, we focus on LoS links as a result of the limited ranges of SUAVs and MAVs dominating the US airspace.

The most throughput intensive link is the ATG payload link (for example
video). In addition to throughput, latency is critical for S\&A video links
which need to be relayed to either the GCS or ATC for decision support. The most important data links in terms of UAV airworthiness are C2 links, which require high reliability and low latency, but lower throughput. A flexible communications system is needed to satisfy these requirements. OFDM can provide high spectral efficiency and flexible data rates, and use more or less spectrum as needed and available. We are thus interested in wideband channel models for ATG links to derive suitable OFDM waveform parameters. 

SUAVs and MAVs typically operate at altitudes of less than 3 km. Commercial applications will predominantly take place in built-up areas. As part of our prior work \cite{Jaber}, we have conducted an exhaustive literature survey on ATG channels (see \cite[Chapter 4]{Jaber} for details). Out of those conducted wideband measurement campaigns and the availability of information, we believe that measurement results available in \cite{Newhall} best describe fading characteristics of SUAV and MAV channels during the en-route phase. Issues in using the results from \cite{Newhall} are, on the one hand, that no information about the Doppler spread is available. On the other hand, the frequency of 2.05 GHz and the position of the GCS at ground level are not typical configurations under which future SUAVs and MAVs will operate. 
We retrieve a power-delay-profile (PDP) prototype defined by the delay vector $\tau=[0\; 33\; 70\; 115\; 175\; 262\; 405\; 682]$ ns and the normalized power vector $P_{dB}=[0\; -8.7\; -9.6\; -11.3\; -13.4\; -15.2\; -17.0\; -20.2]$. The resulting RMS delay spread then equals $\sigma_{t}=87.5$ ns and the maximum observed excess delay spread $\tau_{\max}\approx 1.5$ $\mu$. Our PDP prototype is similar to 3GPP's rural area channel model (for default velocities of $v=\{120,180\}$ km/h and classical Jakes Doppler Spectrum) \cite{3GPP}, which suggests a RMS delay spread of $\sigma_{t}=100$ ns.

%% file: content/waveform.tex
\section{Waveform Management} 
During the en-route phase, different UAVs operate at different maximum mission speeds. Typical MAV and SUAV velocities are 40 and 120 km/h, respectively. It is important to minimize inter-carrier interference (ICI) caused, among others, by the Doppler shift and carrier frequency offsets. We therefore define the following optimization problem to determine an appropriate OFDM subcarrier spacing $\Delta f$ \cite{Das}:
\begin{align}\label{argmax_sc}\Delta f^{*}=&\argmax_{\Delta f_j}\frac{1}{\text{BW}\Big(\frac{1}{\Delta f_j}+N_{CP}T_{s}\Big)}\times\nonumber\\&\sum_{k\in A_{\text{eff}}^{j}}b_{L}(k,\Delta f_j)(1-\text{BER}_{\text{AWGN}}(k,\Delta f_j))\end{align}
subject to\begin{align} 
\label{cond_11}\Delta f_j<&\hat{B}_{c},\\  \label{cond_12}\frac{1}{\Delta f_j}+N_{CP}T_{s}<&\hat{T}_{c}, 
\end{align} where $N_{CP}T_s$ denotes the duration of the OFDM cyclic prefix (CP), $\text{BW}$ is the allocated bandwidth, and $A_{\text{eff}}^{j}=\{0\leq q\leq N-1|q\notin\text{virtual carrier}\}$ represents the set of subcarriers except for virtual subcarriers (guard band). 

\begin{figure}[H]
\centering
\includegraphics[width=2.90in]{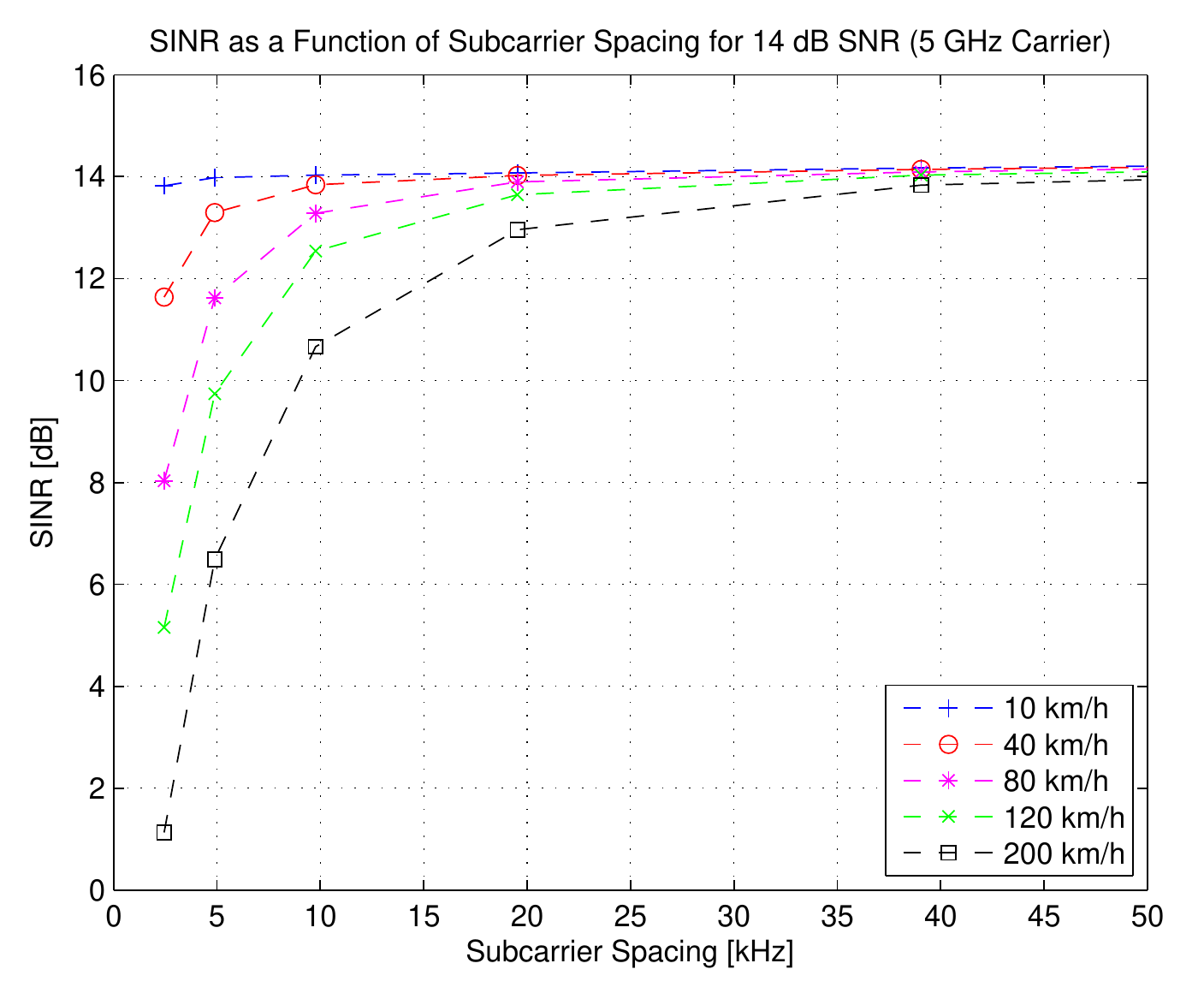}
\caption{\label{fig:SINR_5GHz} \small SINR as a function of the subcarrier spacing at 14 dB SNR.}
\end{figure}

$A_{\text{eff}}^{j}$ is fixed such that $|A_{\text{eff}}^{j}|\Delta f_j\leq \text{BW}_{\alpha}$, where $\text{BW}_{\alpha}=\alpha\cdot\text{BW}$
for $0\leq\alpha\leq 1$. The FFT-size $N$ is chosen to be the "closest" radix-2 FFT size to the number $\lfloor\text{BW}_{\alpha}/\Delta f_j\rfloor$. For $M$-QAM, we define $b_{L}(k,\Delta f_j)$ as the associated per subcarrier QAM bit load according to \cite{Chung} and $\text{BER}_{\text{AWGN}}(k,\Delta f_j)$ as the bit-error-rate (BER) in an AWGN channel at subcarrier $k$ for subcarrier spacing $\Delta f_j$. Above problem optimizes the spectral efficiency subject to constraints \eqref{cond_11} and \eqref{cond_12} which guarantee that the subcarrier spacing and the OFDM symbol duration do not exceed the coherence bandwidth $\hat{B}_c$ and the coherence time $\hat{T}_c$, respectively.

\begin{figure}[h]
\centering
\includegraphics[width=2.5in]{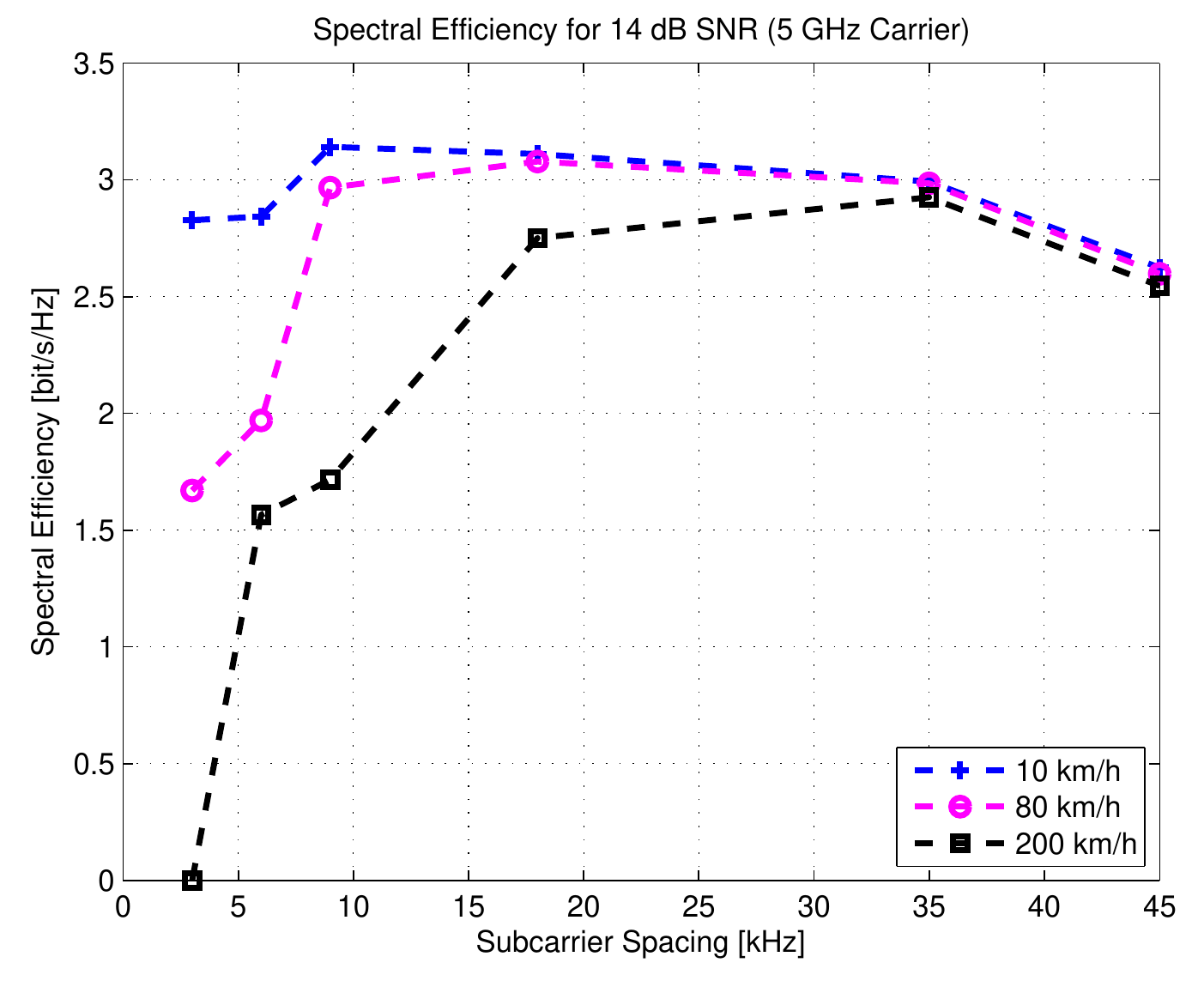}
\caption{\label{fig:spec_eff_5GHz}\small Spectral efficiency as a function of the subcarrier spacing at 14 dB SNR for a 5 GHz carrier.}
\end{figure}

\begin{figure}[h]
\centering
\includegraphics[width=2.5in]{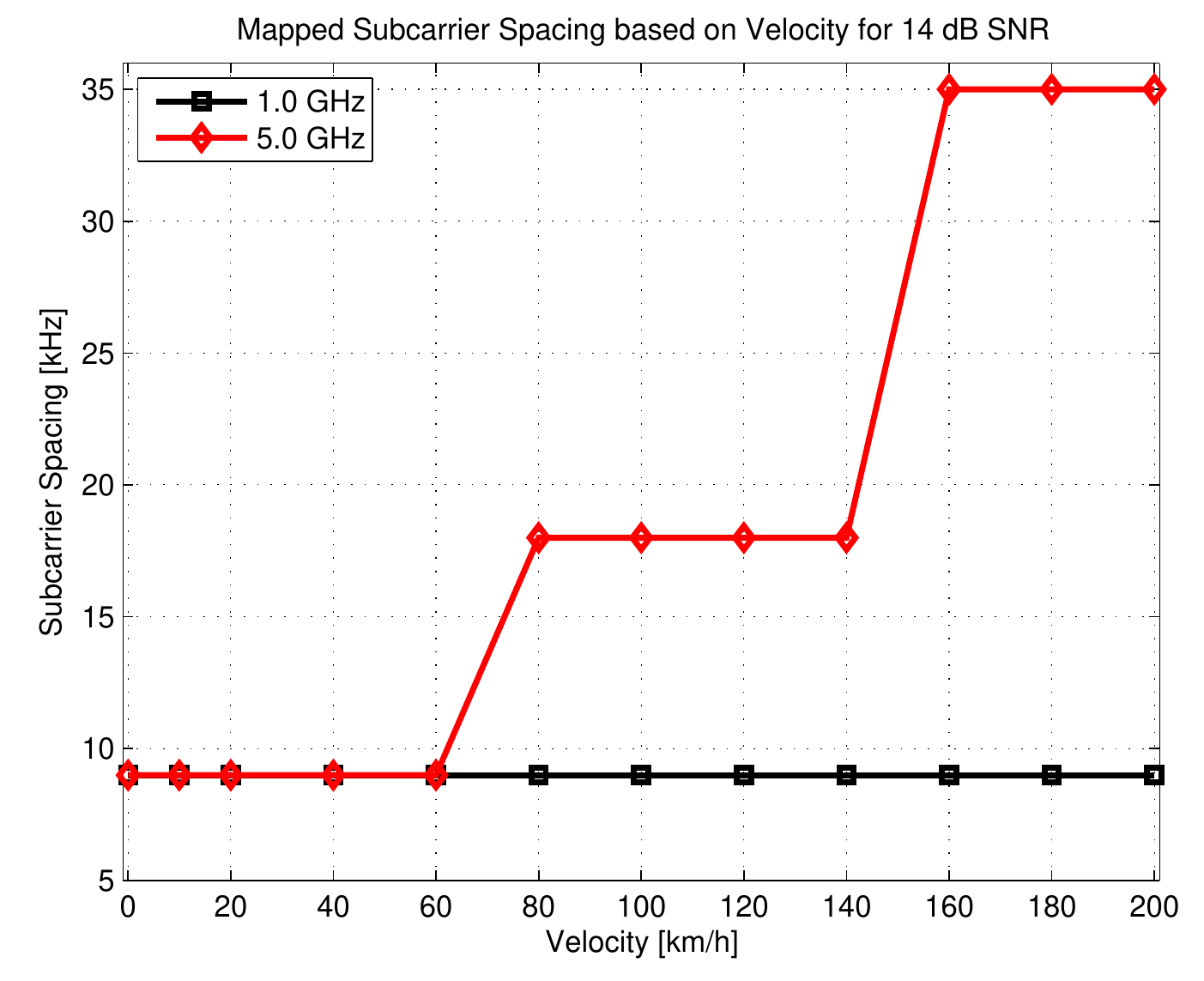}
\caption{\label{fig:map_vel_sc}\small Optimal subcarrier spacing as a function of velocity at 14 dB SNR for 1 and 5 GHz carriers.}
\end{figure}

For our simulations, we fix the available channel bandwidth and the effective signal bandwidth to $\text{BW}=5$ MHz and $\text{BW}_{\alpha}\approx 4.5$ MHz. The PDP (using Jake's Doppler Spectrum) as defined in the previous section is used and a CP length of $N_{CP}T_{s}=2$ $\mu$s chosen to account for the maximum delay spread of $\tau_{\max}\approx 1.5$ $\mu$s with some extra margin.  Based on link budget calculations (cf. \cite[Section 4.1]{Jaber}), the minimum SNR required at the GCS receiver is set to 14 dB. Figure \ref{fig:SINR_5GHz} shows the average \emph{SINR} per subcarrier at carrier frequency of $f_{c}=5$ GHz. As expected, low SINRs are observed for relatively low $\Delta f$ at high velocities. 
Solving the aforementioned optimization problem for the finite set of potential subcarrier bandwidths $\Delta f_j=\{3.0, 6.0, 9.0, 18.0, 35.0, 45.0\}$ kHz produces Fig. \ref{fig:spec_eff_5GHz}. For a given velocity, there is an optimal value for $\Delta f_j$. 
Figure \ref{fig:map_vel_sc} provides this optimal mapping of velocity to subcarrier spacing at 1 and 5 GHz carrier frequencies. For 1 GHz the solution to the problem is $\Delta f_{\text{1 GHz}}^{*}=9\text{ kHz}$. 
Analogously, for 5 GHz we obtain 
\begin{equation}\label{eq:opt_sol_sc_5GHz} \Delta f_{\text{5 GHz}}^{*}=\begin{cases}9\text{ kHz},&\quad\text{if }v\leq 60\text{ km/h}\\18\text{ kHz},&\quad\text{if }80\leq v\leq 140\text{ km/h}\\35\text{ kHz},&\quad\text{if }160\leq v\leq 200\text{ km/h.}\end{cases}\end{equation}
These two frequencies were chosen because L-band (950--1450 MHz) and C-band (4--8 GHz) spectrum are likely to be allocated for UAV communications. 
Note that the above results assume a constant PDP. A frequency-dependent PDP would provide higher accuracy. Nevertheless, our solution can be used as a good approximation for designing and managing ATG OFDM waveforms. 
Based on our results, a $\Delta f$ of 9 kHz for the L-band and 35 kHz for the C-band provide best results for high velocities. For comparison, Eurocontrol's L-band waveform design of L-DACS1 uses a subcarrier spacing of 9.76 kHz. We suggest, in agreement with \cite{Jain12}, using the time division duplex (TDD) mode for multiplexing UL and DL transmissions as it allows asymmetric and variable data rates. UL and DL control information need to be updated at a minimum rate of 20 Hz (or 50 ms) \cite{ReportITU}. We therefore suggest the implementation of a \emph{fixed} UL-to-DL traffic ratio to simplify network synchronization and interference management. Another advantage of TDD over FDD is a faster and simpler channel estimation. TDD is also motivated by the lack of paired spectrum availability (particularly in the L-band) \cite{Jain12}. Note that TDD requires a UL/DL guard time in addition to the CP. This guard time is attributed to the propagation delay and the time needed for the transceiver to switch from receive to transmit mode. This mode is reasonable for SUAVs and MAVs because of their relatively short communications ranges. 

%% file: content/spectrum_sharing.tex
\section{Spectrum Management Discussion} 

RF spectrum may be allocated \emph{directly} to UASs and/or \emph{indirectly} through terrestrial wireless service providers. Network providers can expand the existing infrastructure in support of UAS communications. We envision an integrated solution, where a wireless network provider leases infrastructure resources to, for example, terrestrial cellular service providers and UASs; similarly, an radio resource provider manages access to RF spectrum \cite{vtm12}. 

Spectrum sharing is a natural extension of OFDM-based systems, such as LTE. It is considered an integral part of 4.5G (LTE-Unlicensed) and next generation 5G systems. Spectrum sharing technology is particularly promising here for non-CNPC data since enough dedicated spectrum will not be available. For the sake of airworthiness of UAVs, CNPC data may initially not be carried over shared spectrum. In either case, but especially for DSA, cell coordination becomes crucial to avoid significant interference. 

A UAS communications cell takes a 3D shape, such as a cone, sphere or cylinder, and may overlay with other cells. Different types of cells can be defined for different flight parameters (height, speed, etc.) and services (type of data and communications patterns) to ensure that critical communications services receive the highest protection. The difficulty lies in the large footprint of ATG signals due to elevation and low propagation losses. Static exclusion zones, as proposed for terrestrial spectrum sharing, then become an inefficient spectrum management solution. Instead we propose a centralized, database-empowered spectrum access system (SAS) that facilitates controlled access to shared spectrum as a function of actual UAV positions and frequently-updated radio environment maps. Spectrum allocations can be done on the basis of priorities, fairness, or more complex policies and rules. The inherent latency associated with requesting and receiving spectrum access grants can be improved by careful infrastructure planning, replication of control elements (databases, SASs), dedicated control links, and by combining spectrum management with UAV route planning, among others.

\section{Conclusions}

This paper has analyzed UAV growth and RF spectrum requirements, derived suitable OFDM waveform parameters, and discussed spectrum sharing as a promising solution for meeting future UAS communications demands. The research opportunities are abundant. Network coexistence can be achieved by carefully combining flexible hardware and software technology with efficient and secure protocols, procedures and enforcement mechanisms. Scalable solutions are needed for (1) establishing safety-critical CNPC links and (2) creating mission-critical payload data communications opportunities to satisfy future UAV missions.